\documentclass[sigconf]{acmart} 
\usepackage{multirow}
\usepackage{paralist}
\AtBeginDocument{%
  }

\setcopyright{acmlicensed}
\copyrightyear{2018}
\acmYear{2018}
\acmDOI{XXXXXXX.XXXXXXX}
\acmConference[Conference acronym 'XX]{Make sure to enter the correct
  conference title from your rights confirmation email}{June 03--05,
  2018}{Woodstock, NY}
\acmISBN{978-1-4503-XXXX-X/2018/06}

\begin{document}

\title[Strategic Adaptation Under Contextual Change]{Strategic Adaptation Under Contextual Change: Insights from a Dyadic Negotiation Testbed for AI Coaching Technologies}


\author{Mobasshira Akter Urmi}
\affiliation{%
  \institution{Bellini College of Artificial Intelligence, Cybersecurity, and Computing, University of South Florida}
  \city{Tampa, Florida}
  \country{USA}}
\email{mobasshiraakter@usf.edu}

\author{Raiyan Abdul Baten}
\affiliation{%
  \institution{Bellini College of Artificial Intelligence, Cybersecurity, and Computing, University of South Florida}
  \city{Tampa, Florida}
  \country{USA}
}
\email{rbaten@usf.edu}






\renewcommand{\shortauthors}{Urmi, et al.}

\begin{abstract}
Strategic adaptation—the ability to adjust interaction behavior in response to changing constraints and leverage—is a central goal of negotiation training and an emerging target for AI coaching systems. However, adaptation is difficult to evaluate because adaptation-relevant moments arise unpredictably in typical tasks. We study a reusable dyadic negotiation testbed that employs a controlled midstream change in one party’s outside alternative as a repeatable perturbation to stress-test adaptation. In a six-round chat-based negotiation study ($N=100$), the perturbation reliably reorganized interaction dynamics: transitions between integrative (cooperative) and distributive (positional) behaviors declined, behavioral diversity narrowed, and interactions drifted toward more distributive tactics. Critically, this distributive drift predicted worse relational experience net of objective outcomes, and adaptation patterns were path dependent on prior behavior. These results establish a methodological bridge for evaluating and comparing AI coaching systems on strategic adaptation as a process and identify failure modes and design targets for adaptive interaction support.
\end{abstract}



\received{20 February 2007}
\received[revised]{12 March 2009}
\received[accepted]{5 June 2009}

\maketitle
\section{Introduction}
Many consequential human interactions unfold under changing conditions~\cite{olekalns2003think}. As alternatives, constraints, and leverage evolve, success may depend on how well people adjust their behavior over the course of interaction~\cite{wheeler2002negotiation,druckman2013punctuated,kray2007implicit,olekalns2000understanding}. In negotiation, this adjustment often involves shifting between more integrative (cooperative) behaviors and more distributive (positional, assertive) behaviors as circumstances change. This capacity to purposefully adjust one’s approach in response to evolving conditions—often referred to as \emph{strategic adaptability}—is a central target of negotiation training programs and a growing aspiration for interactive training technologies~\cite{stevens2018using,heunis2025strategic,Shaikh2024Rehearsal,chapman2017proposed}. Yet despite the prominence of this skill, it remains difficult to rigorously evaluate whether a training method or interactive system actually improves people’s capacity to adapt under changing conditions.

A key challenge lies in the \emph{evaluation} settings used to study and train adaptation. Most classroom and laboratory negotiation tasks assume that the rules of the game remain static throughout, limiting opportunities for learners to recognize, process, and respond to changing constraints~\cite{baber2022confirming,brett2016negotiation}. While “turning points” that demand adaptation may still emerge organically through conversation (e.g., impasse or conflict escalation), such moments occur unpredictably and unevenly across dyads. As a result, adaptation-related failures—such as becoming overly positional after a power shift or failing to repair a strained interaction—may not reliably occur or cannot be systematically compared. This poses a challenge for HCI research, where evaluating and comparing interactive training systems requires dynamic tasks that are repeatable, instrumented, and capable of reliably surfacing meaningful contextual change.

\begin{figure*}[t]
  \centering
  \includegraphics[width=\linewidth]{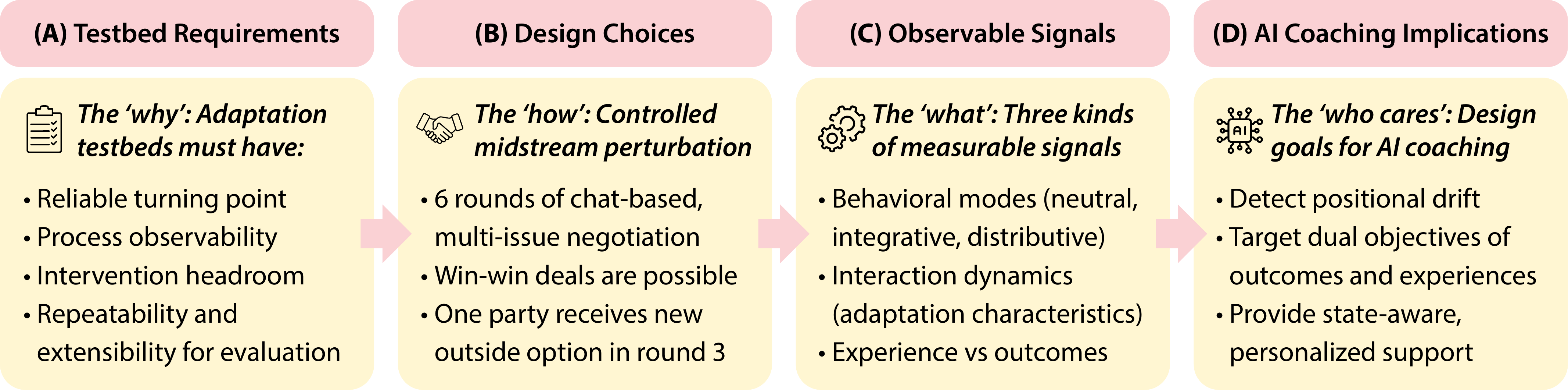}
  \caption{Throughline from adaptation testbed requirements to design choices, observable signals, and AI coaching implications.}
  \Description{A four-panel landscape schematic illustrating the paper's throughline: (A) testbed requirements for studying strategic adaptation (reliable turning point, process observability, intervention headroom, repeatability), (B) the implemented design (a six-round chat-based multi-issue negotiation with a controlled midstream change to one party's outside option), (C) the resulting measurable signals (behavioral modes, interaction dynamics, and experience versus outcomes), and (D) implications for AI coaching (detecting positional drift, optimizing dual objectives, and state-aware personalization).}\label{throughline}
\end{figure*}

In this paper, we study a simple but powerful design move to address this challenge: introducing a controlled midstream change to a negotiator’s outside alternative at a fixed point in an otherwise stable, multi-issue negotiation. An outside alternative refers to what a person could do if the current negotiation fails—for example, accepting an alternative job offer. Rather than aiming for full ecological realism, we treat this change as a controlled exogenous perturbation: a reliable contextual shift designed to stress-test strategic adaptation. 

We examine this design in a chat-based HR–job candidate negotiation conducted over six rounds. By instrumenting interaction processes before and after the turning point, and pairing behavioral measures with both objective outcomes and subjective experience, we ask two questions: (1) Does a controlled midstream perturbation reliably reorganize negotiation behavior in measurable ways? and (2) What patterns of adaptation emerge that matter for experience, not just outcomes? Our findings show that the perturbation consistently reorganizes interaction trajectories: transitions between integrative and distributive behaviors declined, behavioral diversity narrowed, and interactions drifted toward more distributive tactics. Crucially, such distributive drift predicts worse subjective experience even when objective outcomes are held constant. We further find that adaptation is path dependent on pre-change behavior, suggesting that effective support must be state-aware.

This work contributes (1) a reusable, instrumented negotiation testbed for controlled stress-testing of strategic adaptation, and (2) empirically grounded insights into adaptation breakdowns and recovery challenges that inform the design and evaluation of AI coaching systems. More broadly, it demonstrates how HCI-oriented evaluation can bridge theory about human adaptation with the practical requirements of interactive support for dynamic, real-world skills.

\section{Related Work}
Negotiation research emphasizes that interaction outcomes depend not only on the issues under discussion, but also on what each party can do if agreement is not reached—an \emph{outside alternative} that shapes leverage (often formalized as the BATNA: Best Alternative to a Negotiated Agreement)~\cite{wheeler2002negotiation}. Changes in outside alternatives can therefore shift the incentives and behavior of negotiators. At the same time, negotiation is increasingly understood as a temporally unfolding process~\cite{olekalns2003phases}, often punctuated by turning points that reorganize interaction trajectories~\cite{druckman2001turning,baber2022confirming,nadler2003learning,o2001distributive,rubin2013social}. While this literature offers rich conceptual accounts of leverage and phases, most empirical studies rely on static tasks where contextual change arises only organically and unevenly, making adaptation difficult to observe, compare, and evaluate systematically. A few negotiation training tasks incorporate dynamics (e.g., New Recruit~\cite{neale_newrecruit_recruiter,holtom2010using}), but remain underspecified for use as reusable, instrumented evaluation testbeds.

A related body of work frames effective negotiation as requiring \emph{strategic adaptability}: the ability to purposefully adjust one’s approach in response to evolving conditions and the unfolding interaction~\cite{martin2020negotiation,hawes2014recognizing,smolinski2020search}. In mixed-motive settings, this often entails shifting between more integrative, cooperative behaviors aimed at joint value creation and relationship maintenance~\cite{beersma2002integrative,pruitt1993negotiation,elfenbein2008some,smithey2004smart}, and more distributive, assertive behaviors aimed at claiming value when leverage or constraints demand it~\cite{gelfand2011psychology,de2000influence,sato2023teaching}. Although this distinction is well established and widely taught, it remains difficult to translate into actionable support for interactive training systems. Advice to “be adaptive” or “switch strategies” is not operational unless adaptation needs can be detected from observable signals, appropriate responses can be scaffolded in the moment, and efficacy can be evaluated as a process rather than inferred from outcomes alone.

Recent work on AI support for negotiation skill development has shown promise in tactical guidance and knowledge acquisition~\cite{ma2025chatgpt,falcao2024making,shea2024ace,cummins2024friend,duddu2025does,baten2021technology}. However, these systems are typically evaluated in static-context tasks, where the need for strategic adaptation is limited or emergent rather than systematically induced. This leaves an important gap in HCI research: the lack of reusable interaction paradigms that systematically introduce contextual change and allow adaptation to be evaluated as an interaction process.

\section{Methods}
\subsection{Testbed design rationale and payoff construction}
This work engages with a recurring challenge in HCI: how to evaluate interactive systems intended to support complex human skills under dynamic conditions. Prior research has emphasized that controlled laboratory tasks are valuable for making critical phenomena observable, comparable, and diagnosable across participants and interventions~\cite{greenberg2008usability,rogers2012hci}. Synthesizing this perspective with work on negotiation processes and human-AI support, we identify four requirements for a testbed suitable for studying and training adaptation, and design a testbed accordingly (Figure~\ref{throughline}).

\textbf{Reliable contextual perturbation.}  
Organically emerging breakdowns or shifts—while ecologically meaningful—are difficult to analyze systematically~\cite{dourish2001action,ackerman2000intellectual}. HCI studies therefore often introduce controlled perturbations (e.g., injected interruptions or simulated errors) to surface adaptation in a repeatable manner~\cite{lam2021effects,adamczyk2004if}. Following this logic, we introduce a midstream change to one party’s outside alternative as a controlled turning point, ensuring that all dyads encounter a comparable contextual shift.

\textbf{Process observability.} HCI research has long argued that understanding interaction requires examining how activity unfolds over time rather than relying on end-state outcomes~\cite{suchman1987plans,hollan2000distributed}. In negotiation, strategic adaptation manifests as shifts in behavioral stance, switching patterns, and behavioral diversity. The testbed is instrumented to capture such longitudinal process measures before and after the turning point. 

\textbf{Intervention headroom and consequential breakdowns.}
Evaluation-oriented HCI work stresses that useful testbeds must surface non-trivial failure modes rather than be tasks where most users perform near ceiling~\cite{greenberg2008usability}. The task is therefore designed to expose tensions between objective performance and relational experience, and tested on whether they leave meaningful headroom for coaching and intervention~\cite{amershi2019guidelines}.

\textbf{Repeatability and extensibility for system evaluation.}  
Finally, HCI research emphasizes reusable paradigms that allow different systems or interventions to be compared under shared conditions~\cite{rogers2012hci,amershi2014power}. Rather than embedding a specific coaching system, we probe a stable interaction structure with clearly defined instrumentation and a fixed perturbation point, enabling future work to layer alternative training methods or AI coaches while preserving comparability.

To instantiate these requirements, we built on the New Recruit task~\cite{neale_newrecruit_recruiter} to design a chat-based, multi-issue negotiation testbed with a carefully constructed payoff space. Participants negotiated over four issues (salary, bonus, stock options, vacation days), each with five discrete levels. Utilities were additive but asymmetrically structured across roles, intentionally mixing issues with stronger distributive (zero-sum) properties and issues that afford integrative (win-win) trade-offs. We enumerated all possible $5^4$ outcomes to validate the payoff landscape, confirm the presence of a non-trivial Pareto frontier, and ensure that mutually beneficial agreements were available but not immediately obvious (see Appendix~\ref{payoff_appendix} for full details).

Outside-option (BATNA) thresholds were selected using this enumeration. Initial thresholds were symmetric for both parties, requiring moderate utility while preserving a large feasible agreement region. At the midpoint of the interaction, one party’s threshold was increased, introducing an exogenous leverage shift without collapsing feasibility. This controlled perturbation reorganizes incentives while maintaining headroom for continued bargaining and adaptation. 

\subsection{Participants and procedure}
The study was approved by our university’s IRB. We recruited $N=100$ participants ($50$ dyads) from Prolific (demographics reported in Appendix~\ref{demogrphics_appendix}). Participants were randomly paired into dyads and assigned to one of two roles: HR representative or job candidate. Each participant received a \$10 participation payment and could earn up to \$5 in performance-based bonus compensation. Dyads interacted via a web-based chat interface over six negotiation rounds (all code and data will be published open-source upon acceptance). After Round~3, one randomly selected member of each dyad received a notification indicating a change in their outside option; the other party was not informed of this change. The negotiation then continued for three additional rounds. If no agreement was reached by the end of Round~6, each party received the payoff associated with their outside option.

\subsection{Measures}
\noindent\textbf{Behavioral processes.} Each participant’s chat contributions were labeled into three broad behavioral modes: integrative (cooperative, joint-problem-solving), distributive (positional, self-advantaging), or neutral. From these labels, we computed (i)~the proportion of behavior in each mode, (ii)~switching frequency between modes, and (iii)~Shannon entropy~\cite{shannon1948mathematical} as a measure of behavioral diversity over time. Behavioral coding followed the strategic adaptability framework proposed by Heunis et al.~\cite{heunis2024strategic}. The first author coded all turns; the second author independently annotated a random 20\% subset, yielding high inter-rater reliability ($\kappa = 0.9$).

\noindent\textbf{Objective outcomes.} Individual scores were computed from the negotiated package using the predefined utility functions for each issue (Appendix~\ref{payoff_appendix}). Joint scores were computed as the sum of individual scores within a dyad. 

\noindent\textbf{Subjective experience.} After completing the negotiation, participants completed the validated Subjective Value Inventory (SVI), capturing global subjective value and relational experience (rapport)~\cite{curhan2006people,curhan2010objective}.

\section{Findings}
\subsection{The turning point reliably reorganizes behavior toward distributive tactics}
We first examine whether the midstream change in outside alternative functions as a reliable turning point in interaction dynamics. For each participant, we compute PRE$\rightarrow$POST deltas in behavioral measures, estimated using linear regression models with dyad-clustered standard errors to account for interdependence within negotiation pairs. Multiple comparisons are addressed using false discovery rate (FDR) correction. All statistical tests are two-sided.

We observe a systematic contraction of behavioral variety following the turning point (Table~\ref{tab:prepost}). Neutral behavior decreases sharply ($\Delta = -0.150$, FDR-adjusted $p_{\text{adj}} < .001$), indicating reduced exploratory or non-committal interaction. Behavioral entropy also declines substantially ($\Delta = -0.280$, $p_{\text{adj}} < .001$), reflecting a narrowing of the behavioral repertoire. At the same time, participants switch less frequently between behavioral modes, both in terms of normalized switching rate ($\Delta$ switches per step $= -0.092$, $p_{\text{adj}} = .003$) and absolute number of switches ($\Delta = -0.855$, $p_{\text{adj}} = .005$). In parallel, the proportion of distributive (positional) behavior increases ($\Delta = +0.101$, $p_{\text{adj}} = .005$). Taken together, these results show that the controlled midstream perturbation reliably reorganizes interaction trajectories: after the turning point, negotiations become less behaviorally flexible and more position-focused.

\begin{table*}[t]
\centering
\caption{PRE$\rightarrow$POST behavioral changes following the midstream alternative shift. Linear regressions with dyad-clustered SEs.}
\label{tab:prepost}
\begin{tabular}{lrrrrrr}
\toprule
\textbf{Metric} & \textbf{Mean $\Delta$} & \textbf{SE} & \textbf{$t$} & \textbf{$p$} & \textbf{$p_{\text{adj}}$} & \textbf{95\% CI} \\
\midrule
Neutral proportion & $-0.150$ & $0.029$ & $-5.10$ & $<.001$ & $.000002$ & [$-0.21$, $-0.09$] \\
Entropy & $-0.280$ & $0.069$ & $-4.06$ & $<.001$ & $.000169$ & [$-0.41$, $-0.14$] \\
Switches per step & $-0.092$ & $0.028$ & $-3.25$ & $.001$ & $.002685$ & [$-0.15$, $-0.04$] \\
Number of switches & $-0.855$ & $0.286$ & $-2.99$ & $.003$ & $.004913$ & [$-1.42$, $-0.29$] \\
Distributive proportion & $+0.101$ & $0.035$ & $2.88$ & $.004$ & $.005483$ & [$+0.03$, $+0.17$] \\
\bottomrule
\end{tabular}
\end{table*}

\subsection{Distributive drift predicts worse experience beyond outcomes}
We next examine whether these behavioral shifts matter for participants’ subjective experience beyond what can be explained by objective outcomes alone. We estimate actor-level regression models predicting post-negotiation subjective value from PRE$\rightarrow$POST behavioral change, controlling for standardized individual score and joint utility. Standard errors are clustered by dyad, and multiple comparisons are corrected using FDR.

Across models, movement toward distributive behavior emerges as a robust negative predictor of experience. Increases in distributive behavior from PRE$\rightarrow$POST predict lower global subjective value ($\beta = -0.62$, $\text{SE} = 0.21$, $t = -2.98$, $p_{\text{adj}} = .025$) and lower rapport ($\beta = -1.01$, $\text{SE} = 0.20$, $t = -5.02$, $p_{\text{adj}} = .021$), even after controlling for objective outcomes. Importantly, adding behavioral change predictors to outcome-only models yields large increases in explained variance ($\Delta R^2 \approx .18$--$.22$ across dimensions), indicating that interaction dynamics explain substantial variance in experience beyond what is captured by individual or joint performance. By contrast, behavioral change does not reliably predict objective outcomes in this sample, strengthening the interpretation that these effects reflect process costs rather than simply ``people who did worse felt worse.''

\subsection{Adaptation is path dependent on pre-turning-point behavior}
Finally, we examine whether adaptation following the turning point depends on participants’ pre-turning-point behavioral baselines. We estimate reactivity models predicting PRE$\rightarrow$POST behavioral change from standardized pre-turning-point behavior, controlling for post-turn verbosity and dyad clustering. Results indicate clear path dependence. Participants who exhibited higher levels of distributive behavior prior to the turning point showed significantly smaller subsequent increases in distributive behavior ($\beta = -0.41$, $\text{SE} = 0.14$, $t = -2.93$, $p = .004$). In contrast, participants who began the negotiation with more integrative behavior tended to shift more sharply toward positional tactics after the turning point ($\beta = 0.38$, $\text{SE} = 0.15$, $t = 2.53$, $p = .013$). These findings indicate that adaptation is not uniform: the same contextual change can elicit qualitatively different behavioral responses depending on where participants start. This path dependence suggests that effective training interventions may need to be state-aware and personalized rather than relying on one-size-fits-all prescriptions for how people should adapt.

\section{Discussion and Conclusion}
This work shows how a controlled midstream perturbation in interaction contexts can reliably surface strategic adaptation dynamics. The results reveal a recurring, teachable failure mode: after a contextual change, interactions tend to drift toward more positional behavior, with degradation in relational experience net of objective outcomes.

From an HCI perspective, the central contribution of this work is a methodological bridge between theories of human adaptation and the evaluation of AI-supported training systems. Strategic adaptation is widely recognized as a critical human skill. Yet it has remained difficult to support with interactive technologies in part because adaptation-relevant moments arise unpredictably in natural settings, making rigorous evaluation challenging. The controlled perturbation paradigm tested here addresses this evaluative gap by providing a repeatable, instrumented stress test for adaptation. Rather than treating adaptation as an abstract competency, the paradigm renders it observable as a \emph{process}, captured through changes in behavioral mode diversity, transition patterns between modes, and conversational stance before and after a known turning point. This makes it possible to evaluate AI coaching systems not only on whether they improve performance, but on whether they help users adapt \emph{well}—maintaining relational quality while responding to change.

More broadly, the insights have implications for the design of human-centered AI systems that aim to improve social and professional skills. Interpersonal skills are increasingly critical in modern workplaces, where employees must navigate dynamic constraints and collaborate across boundaries~\cite{deming2017growing,elliott2017computers,park2022designing,erlei2022s,govers2024ai,Jaidka2024}. Training technologies that claim to teach dynamic adaptation must therefore be evaluated on dual objectives: task success and interaction quality, in line with prior understanding~\cite{curhan2010objective,curhan2006people}. The testbed studied here provides a foundation for developing and evaluating such AI coaching systems, and can be extended to domains such as teamwork under shifting constraints.
 
\subsection{Design implications for interactive training and AI coaching for adaptive contexts}  
The findings suggest several implications for the design of intelligent negotiation training systems for adaptive contexts:
\begin{enumerate}[(i)]
\item \emph{Detect and respond to distributive drift after contextual change.} The turning point reliably induces a shift toward more positional behavior and reduced behavioral diversity. Because such drift predicts lower subjective values, coaching systems should explicitly monitor for such drift and support repair and reopening of cooperative space. This may include prompts that encourage joint problem framing or scaffold integrative proposal generation.

\item \textit{Treat outcomes and experience as dual objectives.}  
Objective performance and subjective experience are not interchangeable. Behavioral change explains substantial variance in experience even after controlling for outcomes. Training and AI coaching systems should therefore track and optimize both dimensions.

\item \emph{Provide state-aware, personalized support.} Adaptation is path dependent: the same contextual change elicits different responses depending on prior behavior. Systems that tailor guidance based on a learner’s pre-change interaction profile may be better positioned to support effective adaptation than one-size-fits-all advice.
\end{enumerate}

\subsection{Limitations and future work}  
This study has several limitations. First, it examines a single negotiation scenario with crowd workers, and the contextual change is deliberately simplified. Future work should test additional domains and populations. Second, behavioral coding relied on coarse categories; richer linguistic or sequential analyses may be illuminative. Third, we do not examine how individual differences (e.g., personality~\cite{wilson2016personality,elfenbein2008some,elfenbein2018relative,steyer2015theory}) interact with dynamic contexts. Finally, this work does not evaluate a specific AI coaching intervention. Instead, it establishes a validated testbed and identifies adaptation-relevant failure modes, which we will use in future work to design and compare intelligent coaching strategies.

\begin{acks}
This research was supported by a faculty startup fund at the University of South Florida. We thank Anisha Hossain Megha for her contributions to developing the user interface of the data collection website.
\end{acks}


\bibliographystyle{ACM-Reference-Format}
\bibliography{references}

\appendix
\section{Supplementary Materials}
\subsection{Demographic information}\label{demogrphics_appendix}

\setcounter{table}{0}
\renewcommand{\thetable}{A\arabic{table}}

\begin{table}[ht]
\centering
\caption{Participant demographics.}
\label{tab:demography}
\small
\setlength{\tabcolsep}{6pt}
\begin{tabular}{llr}
\toprule
\textbf{Variable} & \textbf{Values} & \textbf{Count} \\
\midrule
\multirow{5}{*}{Age range}
  & 18--24 & 4 \\
  & 25--34 & 32 \\
  & 35--44 & 29 \\
  & 45--54 & 25 \\
  & 55--64 & 9 \\
  & 65+ & 1 \\
\midrule
\multirow{7}{*}{Education}
  & Bachelors & 40 \\
  & Masters & 17 \\
  & Associates & 15 \\
  & Some college & 9 \\
  & High school or equivalent & 13 \\
  & Doctoral & 4 \\
  & Professional & 2 \\
\midrule
\multirow{2}{*}{Gender}
  & Female & 50 \\
  & Male & 49 \\
  & Non-binary & 1 \\
\midrule
\multirow{2}{*}{Hispanic origin}
  & Yes & 9 \\
  & No & 91 \\
\bottomrule
\end{tabular}
\end{table}

\subsection{Payoff design validation}\label{payoff_appendix}
The utility functions are listed in Table~\ref{tab:utility} and the levels used in each issue are listed in Table~\ref{tab:issuelevels}. We enumerated all $5^4 = 625$ outcome combinations and computed individual and joint utilities to validate that the task supports mixed-motive bargaining with non-trivial win--win potential. Across the outcome space, HR utilities ranged from 11 to 63 and candidate utilities ranged from 14 to 70, confirming substantial spread for both roles. The payoff space contains a non-trivial Pareto frontier, indicating multiple efficient agreements with distinct distributional trade-offs.

Outside-option thresholds were selected using this enumeration to ensure a controlled leverage shift without collapsing feasibility. The initial thresholds (35 for both parties) require moderate utility while preserving a large feasible agreement region. The updated threshold (50 for one party) introduces an exogenous leverage change mid-negotiation while retaining feasible agreements above both parties' thresholds.

\begin{table}[ht]
\centering
\caption{Utility functions for each negotiation issue ($l \in \{1,\dots,5\}$). Utilities are additive across issues.}
\label{tab:utility}
\small
\setlength{\tabcolsep}{6pt}
\begin{tabular}{lcc}
\toprule
\textbf{Issue} & \textbf{HR utility} & \textbf{Candidate utility} \\
\midrule
Salary & $3(6-l)$ & $6l$ \\
Bonus & $3(6-l)$ & $3l$ \\
Stock options & $4-(6-l)$ & $2l$ \\
Vacation days & $6(6-l)$ & $3l$ \\
\bottomrule
\end{tabular}
\end{table}

\begin{table}[ht]
\centering
\caption{Issue levels and real-world values used in the negotiation task.}
\label{tab:issuelevels}
\small
\setlength{\tabcolsep}{6pt}
\begin{tabular}{lccccc}
\toprule
\textbf{Issue} & \textbf{L1} & \textbf{L2} & \textbf{L3} & \textbf{L4} & \textbf{L5} \\
\midrule
Salary (USD) & \$80{,}000 & \$90{,}000 & \$100{,}000 & \$110{,}000 & \$120{,}000 \\
Bonus (\% of the base) & 0\% & 5\% & 10\% & 15\% & 20\% \\
Stock options (value) & \$0 & \$5{,}000 & \$10{,}000 & \$15{,}000 & \$20{,}000 \\
Vacation days (per year) & 10 & 15 & 20 & 25 & 30 \\
\bottomrule
\end{tabular}
\end{table}

\end{document}